# Entromics – thermodynamics of sequence dependent base incorporation into DNA reveals novel long-distance genome organization

Petr Pančoška[1,3], Zdeněk Morávek[1], Uday Kiran Para[1], Jaroslav Nešetřil[2,3]

[1]University of Pittsburgh, Department of Medicine, U.S.A.
[2]Charles University Prague, School of Mathematics and Physics, Czech Republic
[3]DIMATIA Center, Charles University Prague, Czech Republic

**ABSTRACT.**

Zero mode waveguide technology of next generation sequencing demonstrated sequence-dependence of the enzymatic reaction, incorporating a base into the genomic DNA. We show that these experimental results indicate existence of a previously uncharacterized physical property of DNA, the incorporation reaction chemical potential $\Delta\mu_i$. We use the combination of graph theory and statistical thermodynamics to derive entromic – series of results providing the thermodynamic model of $\Delta\mu_i$ and showing that it is quantitatively characterized as incorporation entropy. We also present formulae for computing $\Delta\mu_i$ from the genome DNA sequence. We then derive important restrictions on DNA properties and genome assembly that follow from thermodynamic properties of $\Delta\mu_i$ and show how these genome assembly restrictions directly lead to evolution of detectable coherences in the incorporation entropy along the entire genome. Examples of entromic applications, demonstrating functional and biological importance are shown.

**INTRODUCTION.** Besides reading the DNA sequence, application of the zero mode waveguide technology (ZMW) to next generation sequencing (1-5) provides a unique insight into the physical properties of genomic DNA. In the ZMW nano-reactor, fluorescence of a single labeled nucleotide is detected only for the time, which a polymerase needs to chemically incorporate a monomer into the transcribed DNA polymer. When recorded in real time through all the steps of a transcription reaction, these fluorescence pulse duration times $\tau$ vary in a DNA-sequence dependent manner for a given template (5) (see **Fig. 1**). Observed incorporation times $\tau$ have been shown to exhibit a non-Gaussian distribution, both in fast and slow reaction kinetic components (5-8). Here we show that this observation indicates a non-stochastic behavior, indicating existence of a previously uncharacterized physical property of DNA. In the first part of this paper, we use the combination of graph theory and statistical thermodynamics to first derive the sequence-dependence of the incorporation reaction chemical potential $\Delta\mu_i$. We show that $\Delta\mu_i$ is quantitatively characterized as incorporation entropy, and represents an important component of the difference between variants in genome DNA. We also show how $\Delta\mu_i$ is computed from the genome DNA sequence. We then derive important restrictions on DNA properties and genome assembly that follow from derivation of $\Delta\mu_i$. These restrictions are necessary physical conditions guaranteeing that the genome becomes a thermodynamically stable molecular medium with sufficient information capacity. Finally, we show how these genome assembly restrictions directly lead to evolution of detectable coherences in the incorporation entropy along the entire genome. We develop algorithms and tools to detect these long-distance coherence networks and in Part 2. show examples of their application, demonstrating their functional biological role.

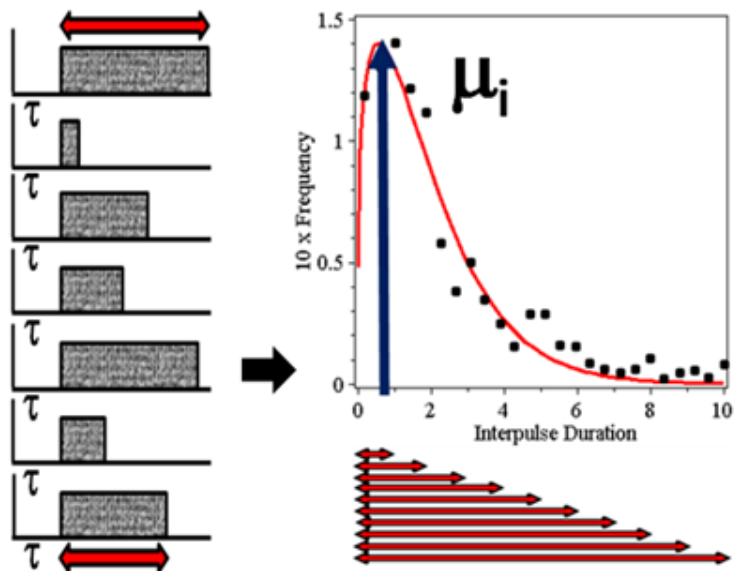

**Figure 1**. *Sequence dependence of the times needed for base incorporation into various DNA position results in the non-Gaussian distribution of incorporation times $\tau$ measured by ZMW experiment.*

## 1. DERIVATION OF ENTROMIC FORMALISM.

**1.1. Contribution of the chemical potential $\Delta\mu_i$ of incorporation to the energy difference between variants in genome sequence.** During enzymatic transcription, the chemical potential $\Delta\mu_i$ of a reaction, incorporating a base into position *i*, is influenced by flanking sequences of *v* bases up- and downstream from *i* (see **Fig. 2**).

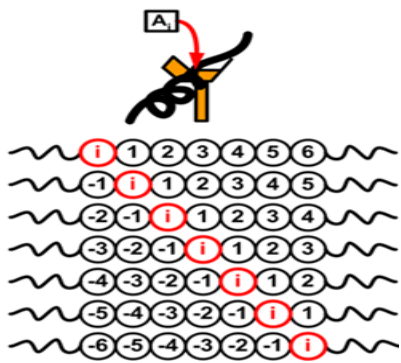

**Figure 2**. *Schematic representation of DNA passage through polymerase in single molecule sequencing experiment.*

Schematically, this dependence can be represented as follows

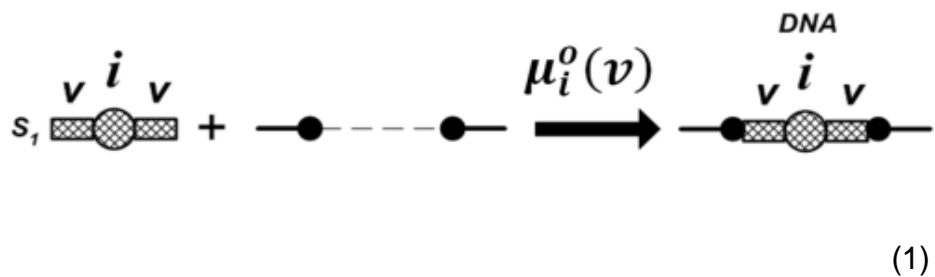

(1)

Replacing the central base by a different nucleotide will lead to an energy difference $\Delta G_i(v)$ between the two DNA variants. The computation of $\Delta G_i(v)$ requires the following thermodynamic cycle

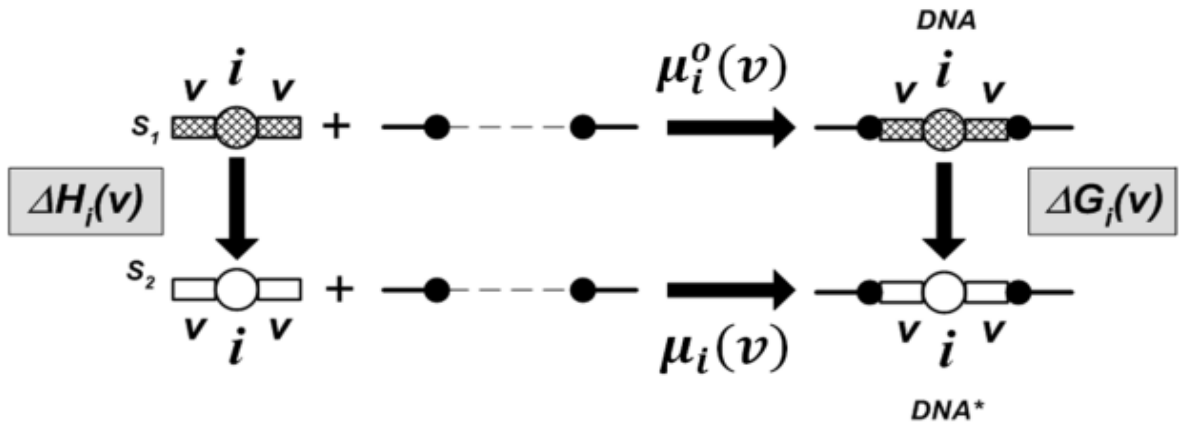

resulting in

$$\Delta G_i(v) = H_i(v) - H_i^o(v) + \mu_i(v) - \mu_i^o(v) = \Delta H_i(v) + \Delta\mu_i(v) \qquad (2)$$

Eq. (2) shows that - in addition to the term $\Delta H_i(v)$, which describes the standard thermal stability difference between two isolated oligomers - the difference $\Delta\mu_i(v) = \mu_i(v) - \mu_i^o(v)$ in the two incorporation reaction energies has to be considered in the complete thermodynamic characterization of the two variant DNA molecules. Thermodynamic models for computing $\Delta H$ from a given DNA sequence are well established (9). To our knowledge, no such formalism exists for using the DNA sequence to compute the incorporation energy contribution $\Delta\mu$. We therefore dedicate the next section to its derivation.

**1.2. Quantitative model for derivation of $\Delta\mu_i$.** We acknowledged that any DNA sequence is processed by a polymerase as a stream of "moving windows". These "moving windows" are DNA segments, which form a finite set [g], defined completely by the transcribed DNA sequence (**Fig.3a**). The thermodynamic properties of these DNA segments, which are relevant for incorporation reaction energy, can be continuous and uniformly distributed or they can exhibit some heterogeneity, which could explain the abnormal sequence dependence of $\tau$. **Fig. 3b** indicates, that any pair of different DNA molecules contains a set of **m** identical "windows" $[g_i]$ (**m** can be zero for completeness), and a set of different segments $[g_i^*]$. Searching for the internal structure of set [**g**] is now more quantitatively reformulated in terms of the difference **d(g,g\*)** between analyzed DNA and an arbitrary selected reference DNA molecule, defining the set [**g\***]. Any equivalence in the sequence-dependent thermodynamic properties of DNA will result in the presence of sets [**g**] with constant **d(g,g\*)**=$\delta$.

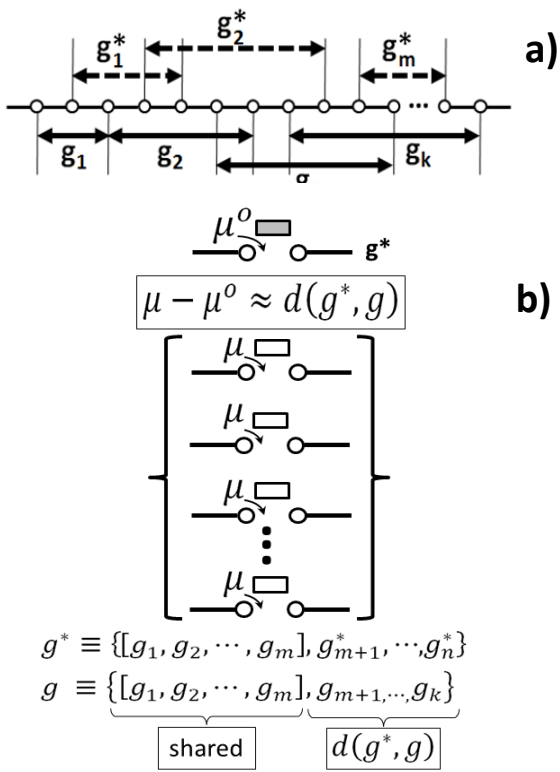

**Figure 3**. *Molecular representation of difference between incorporation entropy of two DNA segments by DNA graph distance **d(g,g\*)**.*
*a) Representation of DNA sequence by series of overlapping segments.*
*b) Partitioning of segment sets into subsets that are shared and different in reference and processed DNA.*

Further progress requires the definition of **d(g,g\*)** in terms of the molecular structure of analyzed DNA. This is made possible by representing the DNA structure by a DNA graph **g**. We use the same symbol to represent DNA segment and its graph, as **g** is a simple cycle graph derived from any given DNA sequence in the conventional way: Vertices of **g** are of four types, representing A,T,C and G nucleotide structures in the actual sequence. Edges in **g** represent the sugar-phosphate bonds and connect vertices in **g** into linear structure, with 5' and 3' end bases being connected to the common end-representing vertex **e** (10,11). The cycle graph representation of DNA structure contains complete information about the segment set [**g**], as can be demonstrated by re-drawing the graph **g** into a template (10,11), or interval graphs (12). The details of DNA structural differences, which can be relevant for incorporation reactions, are now described as a generalized distance **d(g,g\*)** between the graphs **g** of the analyzed DNA and the "central" graph **g\*** of the arbitrary common reference DNA sequence. The tools of combinatorial geometry (13,14), in particular the sampling diversity maximization, are used in the next step of our derivation.

**FACT 1: Maximal sampling diversity characterization of DNA graph sets [g] exists.** We now have a set [g] of DNA graphs g, in which we can define the concentrations of DNA representations, having a specific *d(g,g\*)*. At the same time, *d(g,g\*)* associated with a graph g can be functionally linked to the generalized energy difference between g and g\*. Proposition proved for general graphs in ref. (13) shows that the maximal sampling diversity characterization of a set [g] relatively to [g*] can be found by maximizing the entropy, defined by the probabilities of finding a graph with specific *d(g,g\*)* in the set [g]. Additional constraints on the finiteness of the total number of graphs in [g] and the mean distance *d(g,g\*)* are also considered. The maximal sampling diversity in [g] is characterized by a unique canonical partition function $Z(g^*,\xi)$.

$$Z(g^*,\xi)^{-1} = \sum_{g \in \Gamma} e^{-\xi d(g,g^*)} \qquad (3)$$

In our application, this general result can be interpreted as follows: every DNA graph g represents an "incorporation microstate", which is characterized by the generalized graph distance *d(g,g\*)*. Once we set parameter $\xi$ to be equivalent to the Boltzmann factor, *d(g,g\*)* quantifies the energy difference between these incorporation microstates. Eq. (3) defines just the general form of $Z(g^*,\xi)$, which does not yet provide linkage to the actual DNA sequence. In the next section we show that such linkage can be established through the application of mathematical graph homomorphism theory (15,16).

**FACT 2: Partition function (3) can be computed from the sequence because DNA is a linear molecule.** We stated in section 1.2. that equivalence in the sequence-dependent thermodynamic properties of DNA will result in the presence of molecular subsets in [g] with constant *d(g,g\*)*=δ. As a result, finding such an iso-class of DNA molecules will simplify the partition function $Z(g^*,\xi)^{-1} = \sum_{g \in \Gamma} e^{-\xi d(g,g^*)} = M_i e^{-\beta}$, where $\beta = \delta\xi$ is energy constant and the only variable is $M_i$, i.e. the number of DNA segments having common incorporation energy. In order to compute $M_i$, we first represent the incorporated DNA segment in reaction (1) by a weighted "softcore" graph $H_i$ (see **Fig. 4**). In the physical sense, $H_i$ characterizes a linear DNA polymer as a 4-state system: Edge weights in $H_i$ are transition probabilities, defined for the 18 possible DNA-sequence progressions through molecular monomer states in the linear structure. Using these DNA-molecule defined transition probabilities, $H_i$ can be adapted to represent any actual DNA sequence. One can represent all possible DNA sequences by a single labeled graph $\Gamma$ (see **Fig.5**), which is just a concatenation of corresponding cycles **g**. Softcore graph $H_i$ is defined in order to probe the global properties of a large graph $\Gamma$. In this process, $\Gamma$ is transformed into $\Gamma^*$ (**Fig. 5**) where

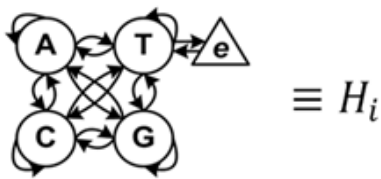

**Figure 4**. *Topology of "softcore" graph $H_i$ for probing global properties of $\Gamma$.*

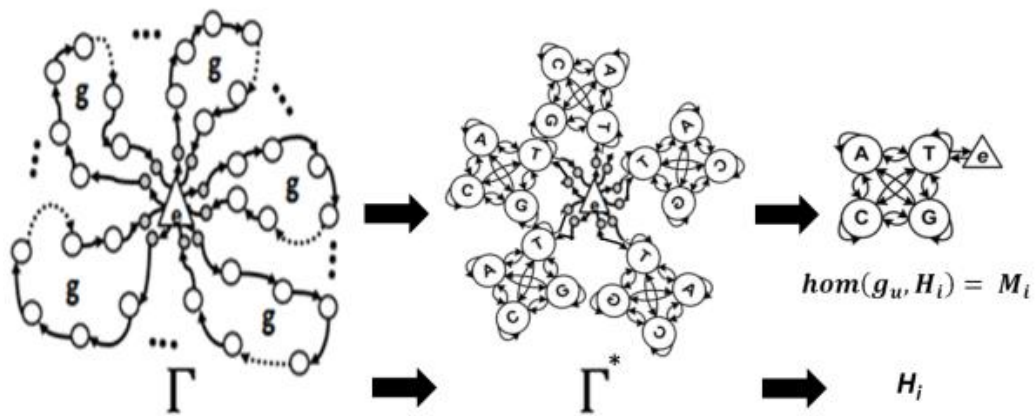

**Figure 5**. *Explanation of relationships between graphs $\Gamma$, $\Gamma^*$ and $H_i$ in homomorphism analysis of DNA sequences.*

.

$g_u$ is unlabeled cycle of length $|g|+1$ and $hom(g_u, H_i)$ is the set of all edge-surjective homomorphisms, which preserve root **e**. It is easy to see that according to mathematical graph homomorphism theory (15,16) the number $M_i$ of these iso-distant sequences is equal to $hom(\Gamma^*, H_i)$. Transition from $\Gamma$ to $\Gamma^*$ is nontrivial. Once we construct a subgraph $H_i$ in $\Gamma^*$ from any known "mother" DNA sequence, it automatically collects all graphs **g** in $\Gamma$, that are Eulerian paths in $H_i$. Subgraphs $H_i$ in $\Gamma^*$ depict the representation of all cycles, rooted at **e**, of length |**g**|+1, which are iso-distant from reference DNA graph **g\***. These subgraphs therefore represent all DNA sequences, which satisfy the condition **d(g,g\*)**=$\delta$. DNA linear structure imposes fundamental simplifying restriction on the topology of the **$H_i$** graph, allowing for exact computation of **$M_i$**. Because of DNA linearity, **$H_i$** is necessarily an Eulerian graph (17). We show in Supplement that, given the sequence-derived weights of $H_i$, our modification of the BEST theorem (17) computes **$M_i$** using the formula (S1).

This result indicates the first essential outcome of our analysis: The linearity of DNA imposes unique and characteristically discrete structure on the set of sequence-dependent properties of the genome. We now know that the genome is necessarily assembled from DNA segments, which are selected from discrete and generally large pools of molecules with identical, and characteristic, energy-related properties with similar sequence-dependence. Because all important structural and functional properties of DNA, including the chemical potential of incorporation reaction $\Delta\mu_i$ are sequence-dependent, this overlooked coherence in DNA properties carries important and novel information. We have shown (10,11) that in daughter DNA sequences, the segments representing the set [**$g_k$**] in **Fig. 3** are related to mother DNA by sub-graph permutations. These **$M_i$** sequences therefore share a common set of "window" motifs [**$g_i$**] and thus **$M_i$** is the sequence-dependent concentration of DNA transcribed with degenerate sequence-dependent properties we were looking for. An important structural consequence of this unique property is that "daughter" DNA sequences are generally not similar to "mother" sequence in the classical sense. This explains why the entromic properties and their relationships, which we derive below, are not observed by standard sequence similarity analyses or in multiple sequence alignments.

**FACT 3: Computation of the position-characteristic incorporation reaction chemical potential $\Delta\mu_i(v)$ from the DNA sequence.** With **$M_i$** determined from DNA sequence of segment 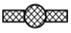 using formula (S1), we can now use the partition function

$Z(g^*,\beta)^{-1} = M_i e^{-\beta}$ for graphs with **d(g,g*)**=δ to derive the chemical potential of the incorporation reaction:

$$\Delta\mu_i(v) = \beta(1/M_i) \qquad (4)$$

This indicates that for a fixed length of DNA segment, the energy cost for the incorporation of a central base of DNA segment decreases with an increasing number of incorporation microstates, in direct dependence on the DNA sequence of the whole segment.

To generalize this result, we need to lift the restriction of eq. (4), which was derived for an arbitrary, but a fixed length of DNA. We now determine the incorporation chemical potential vectors for a series of DNA lengths **v**, varied systematically from 21 to 181 bases, with increments of 2 bases (increase of one base on each side for each step), all centered at position **i**. We normalize these vectors and compute the average $\widetilde{M}_i$ (see Supplement eq. (S3)). The position-characteristic incorporation reaction chemical potential is then computed as

$$\Delta\mu_i = -\beta \ln\left(\frac{1}{\widetilde{M}_i}\right) \qquad (4)$$

$\Delta\mu_i$ is novel, sequence dependent physical property of DNA, capturing the subtle thermodynamic constraints encoded in the arrangement of nucleotides along the genomic DNA. $\Delta\mu_i$ is also an entropy-like property, related primarily to the "topology" of base transition probabilities in the studied DNA. It provides rich and novel information related to non-neutral optimization of the genome function during evolution (18), which is related to important aspects of the gene function. At the same time determination of $\Delta\mu_i$ is concise and computationally simple. In contrast to the thermal stability term $\Delta H_i$ in the expression for $\Delta G_i$ and in accordance with its entropy-like character, $\Delta\mu_i$ is by definition independent of G and C content in the DNA sequence we characterize: The same $\Delta\mu_i$ is computed when bases in the all "daughter" DNA sequences are systematically re-coded (e.g. A to G, T to C, G to A and C to T). To emphasize these unique properties, in further text we will refer to $\Delta\mu_i$ as the **incorporation entropy**.

**FACT 4: Coherence of the incorporation entropy properties is needed for genome stability and high informational capacity.** Analysis of the incorporation entropy (4) for DNA sequences generating extreme "multiplicities" $M_i$ and contrasting the corresponding $\Delta\mu_i$ properties with the general requirements on genome DNA function reveals important general limitation of DNA properties. Importantly, we show below that by analysis of these limitations we can explain the observed distributions of inter-pulse intervals in ZMW experiments. Next we analyze the $\Delta\mu_i$ properties for extreme DNA molecular structures. This essential step provides fundamental physical evidence that selecting <u>multiple</u> DNA segments from the same iso-distant pool during genome assembly is not optional, but rather a necessary and universal genome building principle.

Minimal possible value of $M_i$ is 1. An example of a corresponding DNA sequence is homopolymer (e.g. poly(A)). As log(1) = 0, the "activation energy" for the incorporation of a homopolymeric segment (or periodic repeat stretch) is very high. The other extreme of $\Delta\mu_i$ is found for random DNA sequences, which are associated with the highest $M_i$ values. Eq. (4) indicates that for random sequences with $M_i \to \infty$, an "activation-less" incorporation would be expected Thus, if $\Delta\mu_i$ is considered alone, the general thermodynamic trends would preferentially select for a random genome without repeat regions, which is in direct contrast to biological reality. There is therefore a need for an additional molecular principle to prevent these unwanted trends in assembling the DNA segments into a whole genome.

**FACT 5: The presence of more than one DNA segment with identical $\Delta\mu_i$ in the genome assembly implies Bose-Einstein statistics for the incorporation entropies.**

Through the homomorphism analysis we know that all DNA sequences are found exclusively in "pools", which share molecularly recognizable structural characteristics and, as a direct consequence, unique common incorporation entropies. This implies a simple molecular building principle, sufficient to eliminate the contradictions discussed in Fact 4: The probability of incorporating a single DNA segment with low activation barrier is much higher, compared to the probability of simultaneously incorporating multiple DNA segments with the same incorporation entropy. Similarly, low incorporation probability of a single sequence with high-activation energy into the genome is increased by the requirement that all such segments be incorporated in multiples. This is the same mechanism that is used to remove energy singularities in the Planck characterization of electromagnetic radiation energy distribution.

These qualitative arguments can be made quantitative by adapting the derivation in (19) to our problem (see Supplement for details). We show in previous paragraph that entromic considerations lead to requirement that genome DNA is assembled from "multiplons", which are multiplets of shorter segments sharing the characteristic value of incorporation entropy (4). This also means that multiplon segments are the members of a specific daughter DNA sequence set and associated with the same graph $H_i$. Different permutations of multiplon segments define "assembly microstates" of the genome. The Planck-like distribution function for $\Delta \mu_i$ is derived to characterize the multiplon assembly of genome with the maximal entropy

$$P(\Delta\mu) = A \frac{(\Delta\mu)^a}{\left(e^{\beta(\Delta\mu)(Q-1)} - 1\right)} \tag{5}$$

In eq. (5), the amplitude **A**, the "steepness" of the distribution **a**, mean number **Q** of iso-incorporation entropy segments in the DNA region and effective Boltzmann factor **β** are species-characteristic and genome region specific parameters, which are derived by fitting function (5) to histograms of $\Delta\mu$ values. These histograms are computed from sequenced genome inputs. Note that the (**Q**-1) factor in eq. (5) naturally excludes the presence of just one segment with unique incorporation entropy in the assembled DNA as setting **Q**=1 would result in infinite probability.

There are several ways to validate this novel property of the incorporation entropy coherence in genomic DNA. We show in Results, part 2.1., that eq. 5 fits the experimental distributions of inter-pulse durations from ZMW sequencing experiments.

We also show in part 2.2. that histograms of $\Delta\mu_i$ values, computed from known genome sequences, require fitting by function (5) and that the $P(\Delta\mu)$ parameters are species-specific.

**FACT 6: Location of coherent multiplon segments in the genome can be found by similarity analysis of $\Delta\mu_i(v)$ vectors.** Probing the sequences at distant positions *i* and *j* by softcore graphs $H_i(v)$ and $H_j(v)$ allows the identification of similarities in the sets [$g_k$]$_i$ and [$g_k$]$_j$ irrespectively of the motif order. We can therefore take advantage of having each <u>local</u> position in a genome characterized by the broad <u>non-local</u> DNA sequence context-dependent vector of $\Delta\mu_i(v)$ values. The similarities in DNA motif sets for any two segments from the same iso-incorporation entropy pool lead to the shape similarities in the $\Delta\mu_i(v)$ and $\Delta\mu_j(v)$ vectors along the *v*-coordinate. We quantify these similarities by correlation coefficients $R_{ij} = \int_{v1}^{vk} \Delta\mu_i(v)\, \Delta\mu_j(v) dv$ and visualize them as complete correlation matrices. For better characterization of details, we compute these "overlap integrals" separately for regions with high incorporation entropies, using $\Delta\mu_i(v)$ values

$$R_{ij} = \int_{v1}^{vk} \Delta\mu_i(v)\, \Delta\mu_j(v) dv \qquad (6)$$

and for regions with low incorporation entropies, where we use $1/\Delta\mu_i(v)$ transformation

$$R_{ij}^{-1} = \int_{v1}^{vk} \left(1/\Delta\mu_i(v)\right) \left(1/\Delta\mu_j(v)\right) dv \qquad (7)$$

Computing these matrices allows for direct identification of the genome loci networks with identical or highly similar incorporation entropy characteristics.

**2. RESULTS. ENTROMIC APPLICATIONS AND EXAMPLES.** In this section we show representative examples, demonstrating the main advantage of using entromics' firm physical background for applications to important bioinformatic problems. With entromics, it is possible to formulate a priori (ab initio) non-empirical hypothesis about what is the necessary outcome of an entromic-based algorithm application. This is possible due to the general validity of physics arguments that form the basis of entromics. In classical bioinformatics, the hypothesis formulation is empirical, often descriptive and evidence-based, and thus requires more validation and cross-validation effort in order to establish a successful result.

**2.1. Direct validation of the correspondence between experimental inter-pulse duration distributions and sequence-dependent incorporation entropy distribution** $P(\Delta\mu)$. We extracted experimental inter-pulse duration time values $\tau$ from a ZMW-based, whole human genome sequencing database, available at [http://www.pacbiodevnet.com/]. From $\tau$, measured during the sequencing of 163 DNA segments in the 2,473,644 base long stretch of chromosome 20 we computed 70-bin histograms. We then separated the histogram components, corresponding to the slower (~2 bases/s) and faster (~ 4 bases/s) polymerase processing rates and least squares fitted them by $P(\Delta\mu)$. **Fig. 6** shows typical examples of excellent correspondence between these histograms and the distribution function $P(\Delta\mu)$.

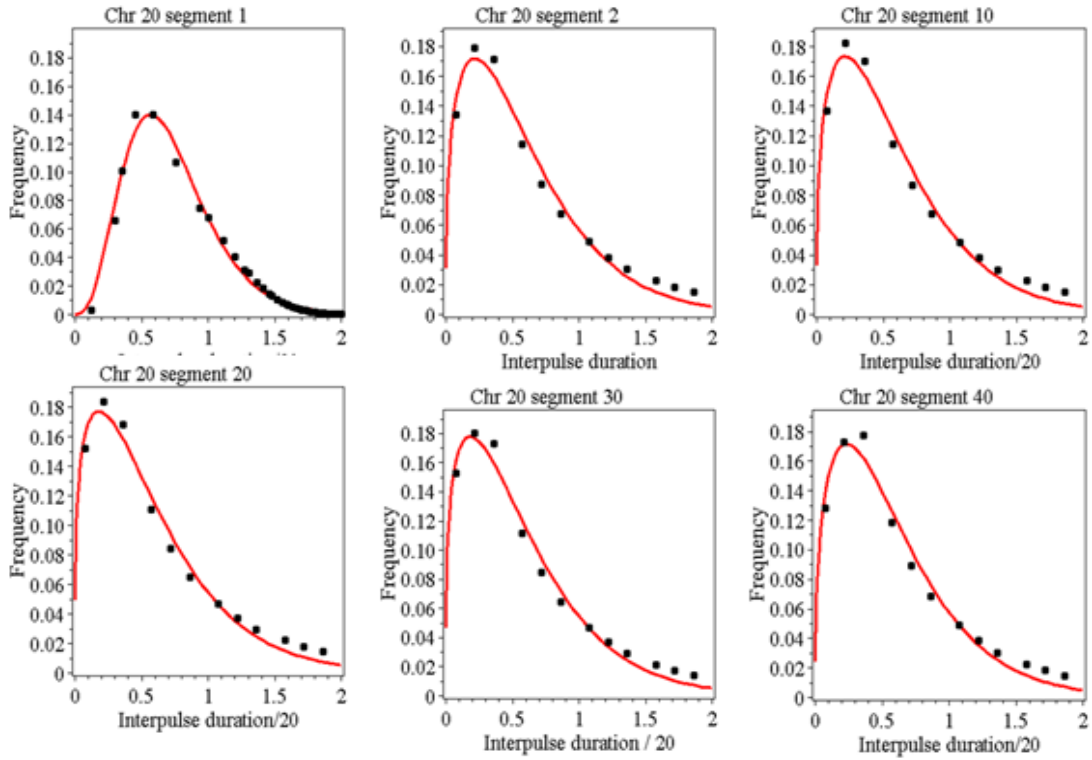

**Figure 6**. *Comparison of experimental distributions of ZMW fluorescence inter-pulse durations sequencing in various regions of human chromosome 20 (circles) with the least squares fit by $P(\Delta\mu)$ (red lines).*

**2.2. Incorporation entropy distributions in genome segments from various species.** Next we demonstrate that the distributions of $\Delta\mu$ values computed directly from a genome sequence satisfy the theoretical prediction by $P(\Delta\mu)$ (eq. (5)). We prepared ~150,000 base long segments, selected consecutively from the genomes of various species. We then computed $\Delta\mu_i$ profiles from the segment DNA sequences (see Supplement for technical details), constructed 70-bin histograms of all $\Delta\mu_i$ values and least-squares fitted them by $P(\Delta\mu)$. **Fig. 7a** shows the representative results for several 150kbase segments of human chromosome 1, demonstrating the exact qualitative as well as quantitative correspondence between the sequence-derived and theoretical distributions (skewed shape, fit significance p<0.0001). **Fig. 7b** shows the species-specificity of the $P(\Delta\mu)$ parameters, obtained by least squares fits of $\Delta\mu_i$ histograms computed from 150 kBase DNA segments of the human chromosome 1, complete genome of *Arabidopsis thaliana,* and three parasites: *Leishmania major*, *Trypanosoma brucei* and *Trichoderma reesei*, respectively.

Results in **Fig. 7b** also demonstrate the potential of entromics to discover novel genome properties. First, the $\Delta\mu_i$ value distributions are not only species specific, but also specific for various position ranges in the genome of one organism. Despite of this intra-species variability, there are significant inter-species shifts in the $\Delta\mu_i$ distribution means. This species specificity opens up the possibility for compact quantitative characterization of important genome properties using the parameters of $P(\Delta\mu)$.

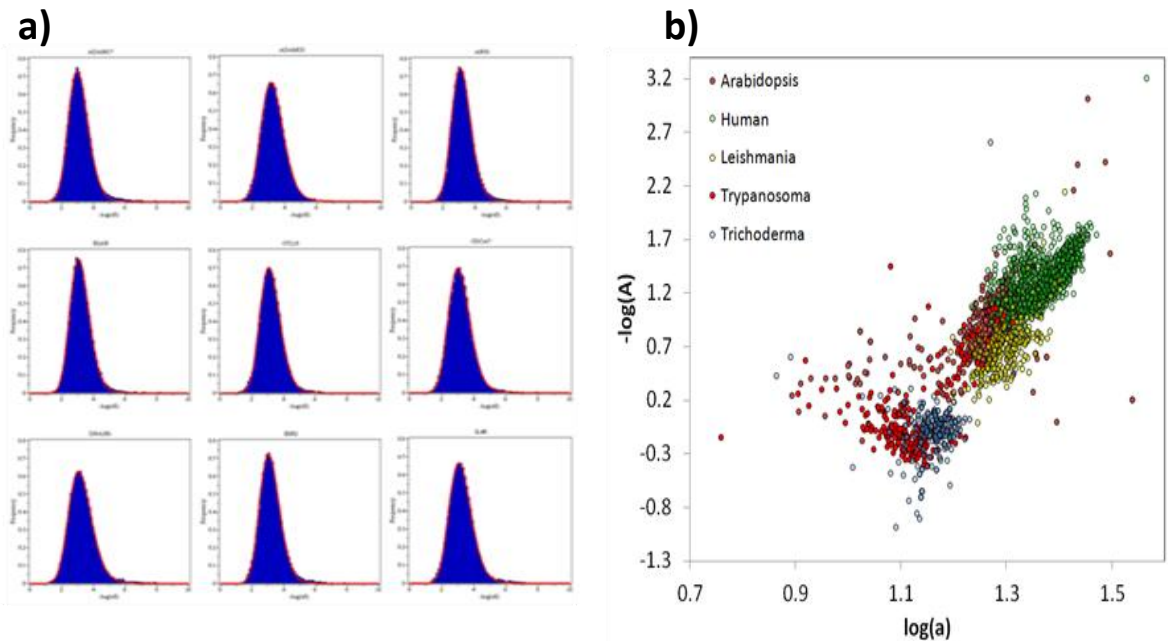

**Figure 7**. *a) Representative distributions of $\Delta\mu_i$ values computed from 150 kbase segments of human chromosome 1 (blue) and their least square fits by $P(\Delta\mu)$ (red lines). b) Inter-species variability of $\Delta\mu_i$ values, represented by parameters A and a of $P(\Delta\mu)$ obtained by least squares fitting of $\Delta\mu_i$ histograms computed from 150 kbase segments in respective genomes.*

We have also studied the phylogenetic dependence of the incorporation energy $\Delta\mu_i$ optimization in the genomes of 14 species, with emphasis on role of codon selection.

Results of this study are being presented in a separate paper, but the pertinent finding was that phylogenetically more complex species have systematically larger differences of the incorporation energy distributions relatively to a reference genomes or exomes with randomized sequences. We mention this result here, because in addition to being evidence for non-neutral evolution across the species, it also contributes to a direct physical explanation of the functional relevance of synonymous mutations. Entromics indicates directly that a local synonymous mutation can lead to energy difference, which becomes distributed to many other genome regions through coherence networking, and has the potential to affect interactions, functionality and dynamic properties of the biological system.

**2.3. Differences between incorporation entropies in the genome and exome.** Due to a detailed understanding of the relationship between DNA sequence properties and $\Delta\mu_i$ we can also predict observable differences in $\Delta\mu_i$ distributions between exome and genome DNA sequences. We reiterate that $\Delta\mu_i$ is directly related to the number of DNA sequences with identical incorporation entropies (see part 1.2.). Any restriction on transitions between the four base states in the genomic sequence naturally reduces this number. Entromics quantitatively captures these restrictions. Amino acid genetic code imposes significant restriction on the entromic properties of exome DNA. We therefore predict that the mean value of $\Delta\mu_i$ distribution, computed from coding DNA sequences in the exome, will be systematically shifted towards lower incorporation entropies, compared to those computed from $\Delta\mu_i$ of the whole chromosome.

**Fig. 8** shows the representative relationship of chromosome vs. exome $\Delta\mu_i$ distributions found in all human chromosomes, confirming the prediction based on entromic principles. This does not mean that non-coding genome parts are not evolutionary optimized through entromic-captured mechanism in the same way as exons are. We have shown (18) that comparisons between $\Delta\mu_i$ distributions computed from actual and randomized intron sequences result in systematic, phylogenetically dependent differences, equivalent to those in coding regions.

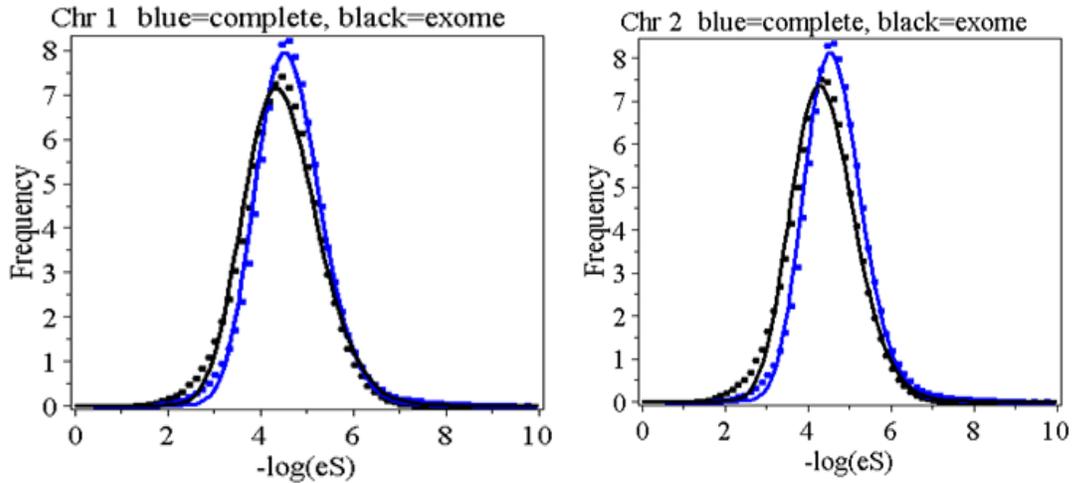

**Figure 8**. *Comparison of $P(\Delta\mu)$ distributions computed in complete chromosomes (blue) and in coding regions (exome) only (black) in chromosomes 1 and 2. Systematic shift of exome distribution means to higher $\Delta\mu_i$ is due to the additional restriction of exon DNA sequences by amino acid coding.*

**2.4. Entromic coherence is species-dependent**. All of the above examples used only the part of the novel entromic characterization of DNA captured in the $\Delta\mu_i$ values. In addition to that, derivation of entromic revealed that there is an additional fundamental long distance networking of genome loci through entromic coherence. Because this second important entromic descriptor is derived from similarity of position-specific vectors $\Delta\mu_i(v)$, the extent of coherence in genomes, captured quantitatively by the correlation matrices [R] and [R$^{-1}$], should be also species-dependent. To validate this
17

biologically essential property, sequences of different species have to be processed together (concatenated) to properly capture the cross-correlation between the entromic properties of the two genomes. To demonstrate this phenomenon, we processed coding DNA sequences from human exome together with those from *Plasmodimum falciparum* and mouse exomes. "Heat map" visualization of computed correlations [R] and [R$^{-1}$], which quantify the extent of incorporation entropy coherences in two extreme networks are shown for *Plasmodium* in **Fig. 9a,b** and for mouse in **Fig. 9 c,d**. Resulting incorporation entropy coherence patterns show clearly different extent of sequence optimization in the three organisms. It is important to re-iterate here, that the clear non-coherence between the human and Plasmodium exomes is not due to the differential GC contents, because both the $\Delta\mu_i$ and $R_{ij}$ values are independent of it. The non-coherence of incorporation entropy between human and parasite exomes is due differences in the level of incorporation entropy optimization, resulting in classically subtle, but entromically highly significant genome context differences, realized through different choices in codon selection. **Figures 9e** and **9f** show that the interspecies differences in the entromic coherence, quantified by $R_{ij}$ are also reflected in the incorporation entropy $\Delta\mu_i$.

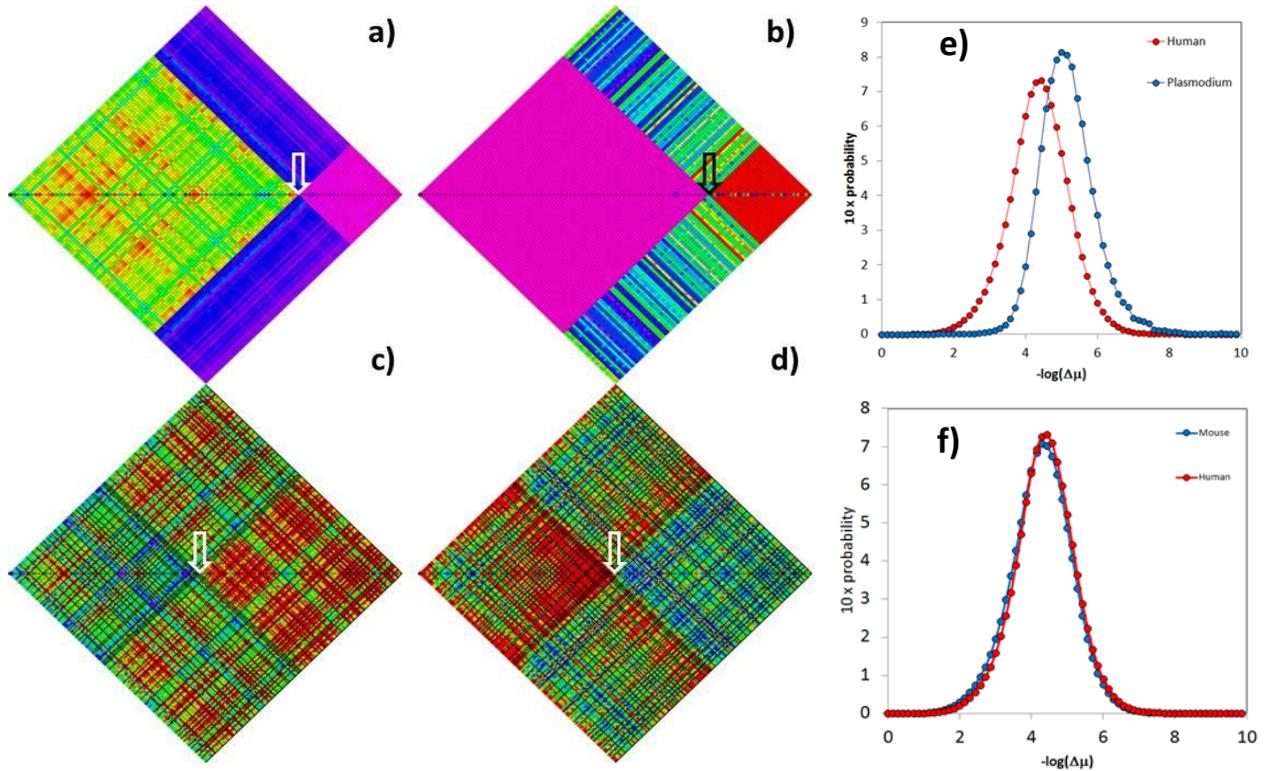

**Figure 9**. *Comparison of entromic coherence matrices in regions with high (top) and low (bottom) incorporation energy $\Delta\mu_i$ computed by exome co-processing for human and Plasmodium (**a,b**) and human and mouse (**c,d**). Human exome results are to the left from the arrow. **$R_{ij}$** intensities increase from magenta to red. (**e,f**) Comparison of $\Delta\mu_i$ distributions for the two species exomes.*

Distribution of $\Delta\mu_i$ for *Plasmodium* is significantly shifted relatively to the $\Delta\mu_i$ distribution in the human exome. While our visualization emphasizes the diversity of human and *Plasmodium* exomes, more detailed analysis of the cross-correlation distributions indicate close matches in the coherences in parts of human and parasite exome, where mutations were associated with the partial resistance to disease in human populations. Human and mouse $\Delta\mu_i$ distributions are very close (see **Fig. 9f**) and the quantification of their differences require closer analysis of entromic coherence patterns.. When the entromic coherence networks are computed with the base resolution, significant differences in the coherence patterns are found even between human and mouse. **Fig. 10** compares the entromic coherence patterns for coding DNA sequences of human and mouse polymerase beta. We first computed the coherence matrices from the two coding DNA sequences and projected the nucleotide-resolution values into protein sequence by averaging the results for the three codon bases. We then represented the 3D profiles of the entromic correlations by contour plots. The thresholds for entromic correlation cut-offs were selected in such a way, that the resulting contours covered the same area fraction (10%) of the *N*x*N* human and mouse correlation planes. In this representation, the MODO algorithm (20) developed for structural alignment of proteins can be used to find an optimal match between the coherence networks in the two polymerase representations. The lines between **Fig. 10a** and **b**, together with the color coding of the polymerase sequence regions indicate segments with optimal matching between the two coherence network patterns in the human and mouse variants. The optimal alignment includes significant shifts, in contrast to practically one-to one correspondence between the amino acid residues in classical sequence alignment showing 90% sequence similarity. The corresponding coherent regions are projected into the x-ray structure of the human polymerase beta in complex with the DNA oligomer, as is shown in **Fig. 10c**. We see that these shifted regions, which do contain the active site, remain in contact with the substrate. Interestingly, this trend of conserving the active site network despite entromic coherence shifts of the structurally important or interaction regions is persistent in many inter-species examples (21). Our understanding of this phenomenon is that the species-specific selection of codons provides information about the set of multiple influences in the whole cell systems biology that have roles in the genomic information processing. The added value from entromic characterization of protein coding sequences is thus the possibility of extracting this otherwise unreachable information. These entromic-exclusive results might be instrumental in better understanding the limitations of animal model experiments and studies.

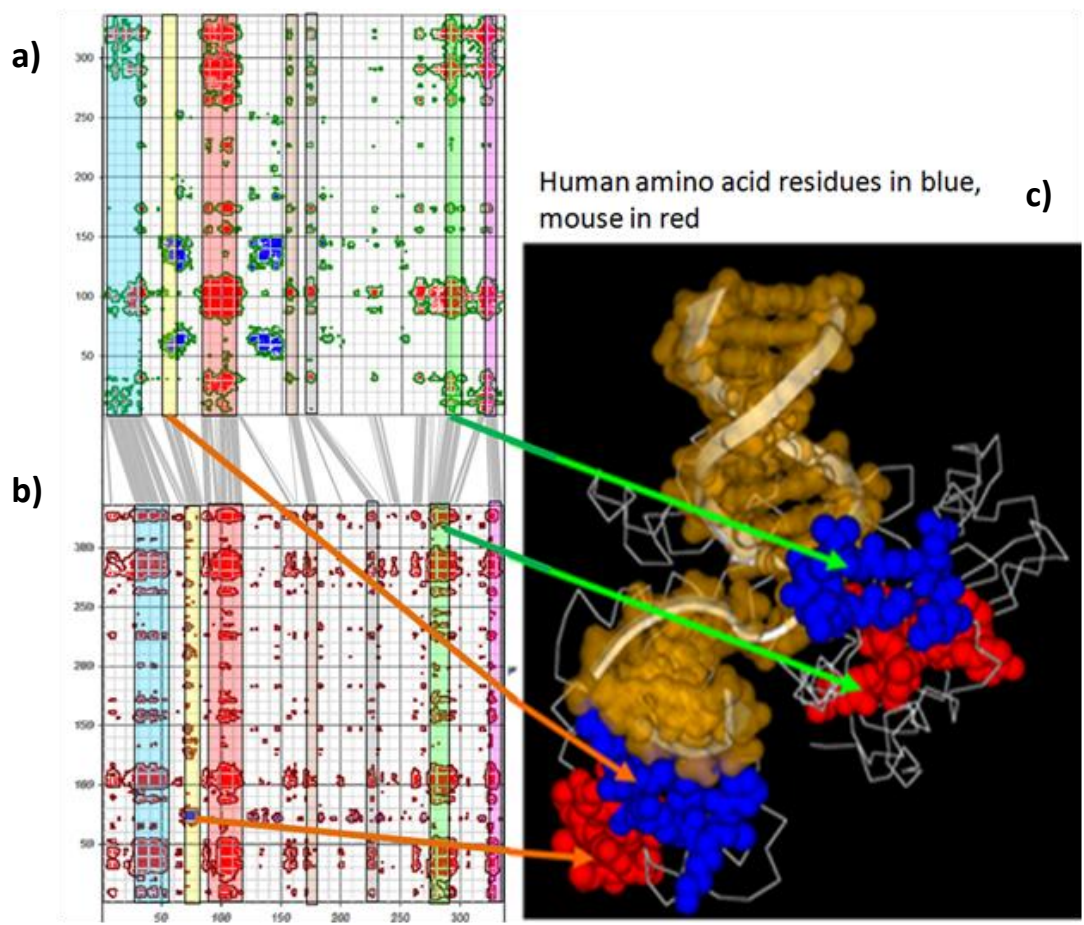

**Figure 10**. *Structural interpretation of shifts in the entromic coherence patterns for human (**a**) and mouse (**b**) polymerase $\beta$. The lines, connecting coherent regions in two species were computed by matrix alignment program (ref. 17). **c)** Projection of aligned coherence regions into human polymerase-substrate structure (3ISB).*

**2.5. Cystic fibrosis – clinical application of entromic.** Severity of cystic fibrosis is critically associated with the presence of CFTR gene variants, such as alternatively spliced CFTR isoforms. Out of 27 exons in the CFTR gene, two (exon 8 and 13) can be skipped during transcription. We were interested if the incorporation entropy can explain this selectivity in CFTR transcription defects. According to the entromic principles, we predict that problematic transcription will more likely occur for DNA segments with a sequence resulting in low incorporation entropy, where the activation barrier for transcription is high. **Fig. 11** shows the computed the $M_i$ (red) and $1/M_i$ (blue) profiles for all exons from the DNA sequence of the CFTR gene. The exons with the lowest entropy gain from incorporation into the gene are those where 1/ $M_i$ profile dominates (see eq. (4)). In **Fig. 11** we show that this was found exclusively in exons 8 and 13, in accordance with their observed alternative splicing.

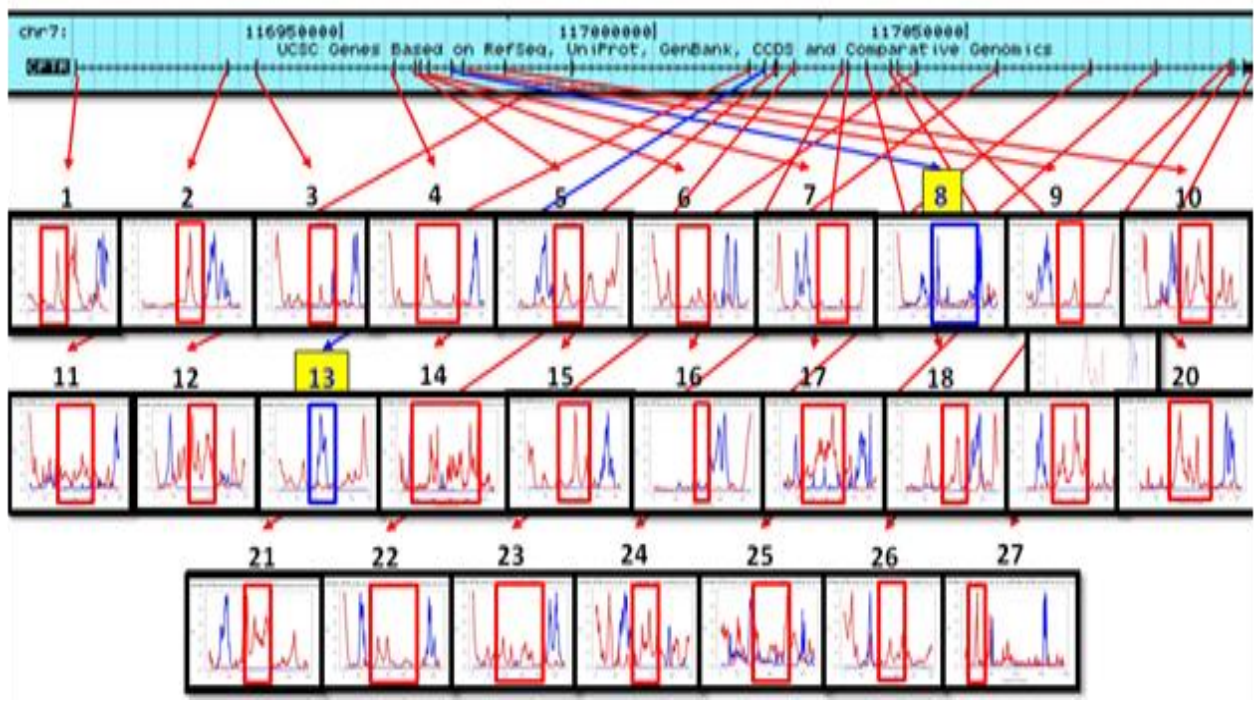

**Figure 11**. *Plots of $M_i$ (red) and $1/M_i$ (blue) profiles in exons of human CFTR gene. Only exons 8 and 13 have dominant $1/M_i$, indicating that corresponding highest incorporation entropy contributes to their specific disease-related splicing.*

Importantly, entromics also indicates that the extent of this splicing will be affected by any mutation in these exons. There is experimental evidence for this sequence-dependence. Pagani et al. (22) systematically analyzed the impact of synonymous mutations on alternative splicing of exon 13 in the CFTR gene. Their experiments have shown that a high proportion of these synonymous mutations (25% of all tested) may induce CFTR exon 13 skipping, consequently reducing normal transcript expression below 25%. In contrast to conventional methods, entromics can quantitatively characterize the difference between the wild type and synonymous mutants by the values of incorporation entropy change. We therefore computed the differences between the incorporation entropy profiles of all exon 13 mutants of CFTR gene and the reference profile for wild type CFTR gene sequence (see **Fig. 12a)**. Because of the non-local characterization of the incorporation entropy, the change of $\Delta\mu_i$, caused by any gene variation results in a difference $\Delta\mu_i$ profile (see **Fig. 12a**). The overall incorporation entropy change is therefore computed as integrated area under these difference $\Delta\mu_i$ profiles. **Fig. 12b** shows that the experimentally determined percent prevalence of the CFTR variant without exon 13 is linearly related to this integrated incorporation entropy differences. This demonstrates the potential that entromic characterization of genome and its variants brings to solving problems that are relevant for quantitative understanding of disease-outcome related genetic factors.

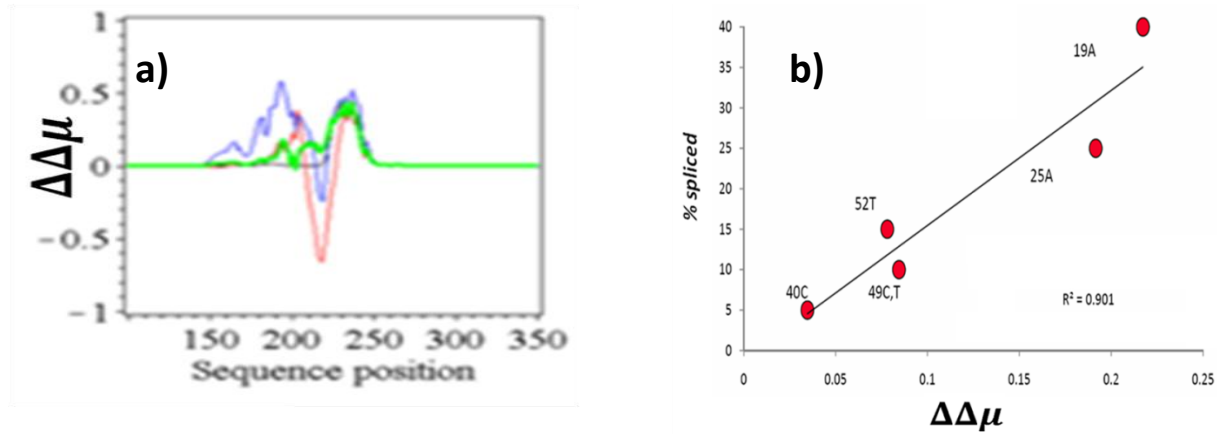

**Figure 12**. *a) Example of difference $\Delta\mu_i - \Delta\mu_{i,wt}$ profiles for all synonymous mutants in one of the CFTR gene exon 13 position, studied in ref. 18. b) Linear relationship between the total $\Delta\mu_i - \Delta\mu_{i,wt}$, computed as integrated areas under the difference curves and the experimental percentage of spliced CFTR variant expression.*

## 3. CONCLUSIONS.

We have shown that sequence-dependence, observed for progression of enzymatic nucleotide incorporations into genomic DNA through single-molecule experiments reveals important internal organization of genome. Combination of basic, but fundamental thermodynamic principles with the homomorphism-based graph-theory analysis of information in the genomic DNA resulted in formalism of entromics. With entromic tools, we matched the sequence-dependent distribution of experimental incorporation times satisfactorily with theoretically derived distribution function $P(\Delta\mu)$. The importance of entromic principles exceeds this primary result. Entromic revealed new internal, species-specific genome organization, which has highly nonlocal character. Eq. (2) indicates that through these nonlocal incorporation entropy coherences a novel compensatory mechanism operates in genomes. Selection of alternative coding or subtle changes in the sequence context in the modified locus as well as in distant loci can compensate for unwanted changes in the thermal stability caused by mutation, polymorphism or necessary evolutionary modifications.

Importantly, entromics also shows that these "footsteps of evolution" in genomes are organized into recognizable interlaced networks of multiplons. Entromic theory provides non-empirical formulae for computing all its state functions. We use these entromic functions in the deterministic algorithm identifying these genome-wide coherence networks. Connection strengths between the respective regions are also computed from the sequenced genomic DNA. This provides general, physical-principle based blueprint of energy-based, and therefore molecularly recognizable, long-range multi-loci, evolutionary optimized inter-relationships in any genome. We have shown (18) that the level of this optimization is highest in human genome and lowest in the phylogenetically simplest organisms. This demonstrates the functional and biological relevance of physical principles underlying the characterization of entromics genome properties and assembly rules. In special cases these physical principles might be directly involved in the biological function or alter that function when sequence is perturbed (e.g. dynamics of processes where transcription rate plays direct role). More importantly, we need to understand these networks as global indicators of interconnection between thermodynamic and functional optimization of system biology processes in a cell. Locations of the coherence-related regions inform us about genome parts with extremely low or extremely high tolerance against variation. Positive or negative changes in entromic properties upon genome variation inform us about the potential of the system to compensate for these alterations through long-distance linkages and their strengths. We can directly compute the differences in the incorporation entropy profiles and coherence strength matrices for original and altered states of genome and use these differences in explanatory models of function. The important advantage of entromic in this type of applications is that these informative and functionally relevant results are computed by equations of entromics physics directly from a single, individual genomic sequence.

Our detailed analysis also explains why the entromic properties were previously undetected. DNA regions in entromic coherence networks do not exhibit significant sequence similarity. At the same time, sequence similarity is special case of entromic coherence (identical DNA sequences are indeed coherent and equivalent in entromic sense). Entromics is therefore not at all in contradiction to existing and successful bioinformatics paradigms. Entromics tools allow opening of novel quantitative characterization of many aspects of genomics that are in existing approaches complex, difficult to handle, hidden or undetectable. Large-scale practical applications of entromics can be efficiently implemented, because theoretical development resulted in analytical instead of approximate or empirical algorithms. Our current platform processes whole genome entromic characterizations, finalized in hours CPU time with results efficiently stored in searchable database, annotated by links to bioinformatics resources and structural and functional visualization tools.

The main advantage of entromic characterization of genomes is that it provides single, deterministic and interpretable network of validated weights, highlighting the functionally dominant genome regions with quantitative characterization of the communication and mutual compensatory capacity between them. This brings new dimension to genomics. We are no more restricted to analyzing just genotype differences. These are in reality just small perturbations of the overall physical "potential" resulting for example in the several orders of magnitude variability in the incorporation energies we now can compute from consensus or individual, personal DNA sequences. The variation of common entromic properties throughout the common 99.9% of genome, shared by all people, are equally if not more important for proper understanding of the biological and disease processes, where genetic factors play an essential role. On the top of these common genetic potential, the differences in the incorporation entropies due to the interpersonal genome variations computed from personal genomes through entromics become much more informative genetic component of personalized medicine.

SUPPLEMENT

**S1. Multiplicity.** Using a modification of well-known results of discrete mathematics (23), the number of sequences in each of DNA segment sets with $d(g,g^*)=\delta$ (an iso-family), the *multiplicity* $M_i$ can be calculated directly from the parent DNA-graph $H_i$, using the following analytical formula:

$$M_i = det(L^*) \frac{\prod_{i=1}^{4}(A_{ii}-1)!}{\prod_{i=1}^{4}(\Delta_{ii})! \, \prod_{i \neq j}^{4}(A_{ij})!} \qquad (S1)$$

Here $A_{ij}$ are matrix elements of the adjacency matrix of graph $H_i$. Terms $\Delta_{ii}$ are calculated as differences between diagonal elements $A_{ii}$ and the row or column sums of off-diagonal elements $A_{ij}$ ($i \neq j$) of graph adjacency matrix $[A]$:

$$\Delta_{ii} = A_{ii} - \sum_{j \neq i}^{5} A_{ij} \qquad (S2)$$

$L^*$ is a Kirchhoff matrix, which is determined by a) selecting a 4x4 submatrix $[L]$ of $[A]$ (i.e. no *e*-rows and columns of $[A]$ are considered); b) changing the off-diagonal terms in $[L]$ to negative numbers and c) replacing the diagonal elements of $[L]$ by $A_{ii} - \Delta_{ii}$. As the determinant of the Kirchhoff matrix is equal to the number of *e*-rooted trees in $H_i$ and the second term enumerates the multiplicity of possible "departures" from each base-defined vertex in polynucleotide continuation, the numerical value $M_i$ is a sensitive function of sequence context.

**S.2. Characterization of position by the vector of incorporation energies for systematically elongated DNA segments.** Figure S1 shows details for computing position-dependent vector of incorporation energies from which the position characteristic mean is computed. Because we need equivalent characteristics for all positions in the sequence, we attach to every processed DNA common "buffer" DNA sequence of length 2001 bases. In this way, both 3' and 5' ends of any actual

processed genomic sequence are considered in the identical context of the same flanking buffer bases (cyclic boundary conditions). We also use the vector $\mu_{1001}^0(v)$ of incorporation reaction chemical potential in the center of this common buffer sequence for systematic normalization of the chemical potential values in the analyzed DNA sequence as is shown in the top part of the Fig. S1.

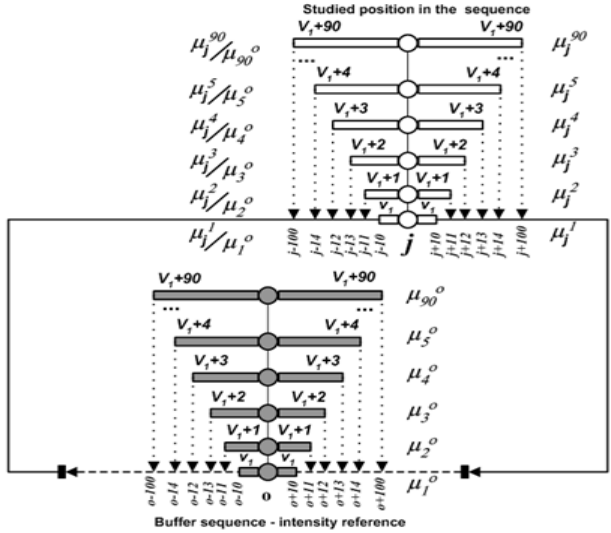

**Figure S1**. *Scheme explaining computation of incorporation entropy vectors from DNA sequence. Series of position-centered segments with systematically increasing length is selected from DNA sequence (top, open rectangles) and processed into incorporation entropy. Simultaneously, the computation is performed for segments of the same length in the center base of always identical "buffer" DNA sequence, which facilitates the common cyclic boundary conditions at the DNA ends (bottom, shaded rectangles). The incorporation entropy values from the reference computation are used for normalization as shown.*

The sequence-dependent $M_i(v)$ values for every segment length (2v+1) are normalized by the corresponding reference $M_{1001}^o(v)$ value. We then compute mean value of these results to obtain length-independent, per-base normalized characteristics of the "daughter" DNA segment concentration at each position $\widetilde{M}_i$,

$$\widetilde{M}_i = \frac{1}{N_v}\sum_1^{N_v} \frac{M_i(v)}{M^o(v)} \tag{S3}$$

from which the value of $\Delta\mu_i$ is computed. This algorithm not only resolves the problem of the optimal characterization of studied DNA ends by the length-dependent vector of sequence-dependent quantities, but implements at the same time the conventional thermodynamic reference standardization of the computed values, which become quantitatively comparable across all DNA sequences, genomes and organisms.

**S.3. Proof of the proposition for maximal sampling diversity in the metric models for random graphs applied to Γ sub-graph sets (13)** Let Γ is a finite set of DNA graphs **g**. Let d(g,g*) is the arbitrary metric on Γ. Define entropy on Γ as $S(p) = -\sum_{g \in \Gamma} p(g) \ln(p(g))$, where $p(g)$ is the probability of finding graph **g** in Γ. The following proposition proves that there exist (Gibbs) distributions of graphs in Γ, which maximizes $S(p)$ and as a consequence, provides maximal sampling diversity in Γ:

Distinguish a reference element g* in G. The probability distribution that maximizes $S(p)$ subject to constraints $\sum_{g \in \Gamma} d(g^*, g) p(g) = \varepsilon$ (conservation of total energy) satisfies $p(g) = Z(g^*, \xi) e^{-\xi d(g^*, g)}$, where $\frac{1}{Z(g^*, \xi)} = \sum_{g \in \Gamma} e^{-\xi d(g^*, g)}$ and $\xi$ is the unique solution to $\frac{d \ln c(g^*, \xi)}{d\xi} = \varepsilon$.

Proof: $S(p) - \Theta - \xi \varepsilon = -\sum_{g \in \Gamma} p(g) \ln p(g) - \Theta \sum_{g \in \Gamma} p(g) - \xi \sum_{g \in \Gamma} d(g^*, g) p(g) =$

$$= \sum_{g \in \Gamma} p(g) \ln \left( \frac{1}{p(g)} e^{-\Theta - \xi d(g^*, g)} \right) \leq \sum_{g \in \Gamma} p(g) \left( -1 + \ln \left( \frac{1}{p(g)} e^{-\Theta - \xi d(g^*, g)} \right) \right) =$$

$$= -1 + \sum_{g \in \Gamma} e^{-\Theta - \xi d(g^*, g)}$$

The inequality reflects the fact that the function values of lnx are below tangent line at x=1, i.e. $\ln(x) \leq x - 1$ for all x>0, with equality if and only if x=1. With the context of $S(p)$, equality occurs if and only if $p(g) = e^{-\Theta - \xi d(g^*, g)}$. This choice maximizes the entropy. Since $\sum_{g \in \Gamma} p(g) = 1$, we obtain $Z(g^*, \xi) = e^{-\Theta} = \frac{1}{[\sum_{g \in \Gamma} e^{-\xi d(g^*, \xi)}]}$. The constraint $\sum_{g \in \Gamma} d(g^*, g) p(g) = \varepsilon$ now results in $\frac{d Z(g^*, \xi)^{-1}}{d\xi} = -\varepsilon Z(g^*, \xi)^{-1}$, implying $\frac{d Z(g^*, \xi)^{-1}}{d\xi} = \varepsilon$.

Monotonicity of Z as function of $\xi$ ensures a unique solution for $\varepsilon$ in the last equation. ∎ This proof is mathematical alternative to Lagrange-multiplier – based solution of equivalent problem in statistical thermodynamics.

**S.4. The classical derivation of Planck distribution of $\Delta\mu_i$ adapted from (19) to DNA system with multiple iso-incorporation entropy segments.** Let $M_i$ be the number of DNA motifs having common incorporation entropy of $\Delta\mu_i$. Let $p_k(g)$ be the probability that there are exactly **k** motifs **g** with this incorporation entropy in assembled DNA. This leads to a constraint $\sum_k k \cdot p_k(g) = M_i$. We substitute $p_k(g) = \frac{n_k}{n}$, where $n_k$ is the number of iso-incorporation entropy motifs in the assembled DNA, consisting of total **n** motifs. With this consequence of DNA linearity, we convert the constraint $\sum_k k \cdot p_k(g) = M_i$ into relationship between "countable" integer quantities $\sum_k k \cdot n_k = M_i n = \Omega_i$. We can now represent the possible DNA assemblies from various multiplets of motifs with degenerate incorporation entropies of k-types, each repeated in the DNA assembly $n_k$ times as linear motif structure 111222333334444455555...... The permutations of this assembly, given the set of $n_k$ values, are "assembly microstates". The total number of these microstates is given by the combination of **n** elements taken $M_i$ at a time, $C_i = \frac{(n+M_i-1)!}{(n-1)!(M_i)!}$. Using Stirling formula, we can rewrite $C_i$ as

$$\frac{1}{n}\log C_i = (1+P_i)\log(1+P_i) - P_i \log P_i, \quad \text{where} \quad P_i = M_i/n.$$

By maximizing $L = (1+P_i)\log(1+P_i) - P_i \log P_i$ under the constraint $\sum_i \Delta\mu_i = \epsilon$ requiring naturally that all $\Delta\mu_i$ values sum into finite total incorporation entropy cost $\epsilon$ and by using Lagrange multiplier method, we find $P_i = \frac{1}{\left(e^{(\beta\Delta\mu_i)}-1\right)}$. Thus, the indistunguishability of incorporation entropies for the $M_i$ DNA motifs associated with the same "softcore" graph $H_i$ lead to their behavior in the assembled genome DNA governed by Bose-Einstein type statistics. Integrating this result to obtain average incorporation entropies for each motif type and considering that DNA motifs have non-zero chemical potential, the result is eq. (5) in the text.

$$P(\Delta\mu) = A \frac{(\Delta\mu)^a}{(e^{\beta(\Delta\mu)(Q-1)} - 1)}$$

(S3)

**FIGURE CAPTIONS**

**Figure 1**. **a)** Schematic representation of DNA passage through polymerase in single molecule sequencing experiment; **b)** sequence dependence of the times needed for base incorporation into various DNA positions; **c)** Non-Gaussian distribution of incorporation times.

**Figure 2**. Thermodynamic cycle for computing the contribution of incorporation energy $\Delta\mu_i(v)$ to difference between two DNA sequences with variant in position i.

**Figure 3.** Molecular representation of difference between incorporation entropy of two DNA segments by DNA graph distance *d(g,g\*)*. **a)** Representation of DNA sequence by series of overlapping segments. **b)** Partitioning of segment sets into subsets that are shared and different in reference and processed DNA.

**Figure 4.** Topology of "softcore" graph **H** for probing global properties of $\Gamma$.

**Figure 5.** Explanation of relationships between graphs $\Gamma$, $\Gamma^*$ and $H_i$ in homomorphism analysis of DNA sequences.

**Figure 6.** Comparison of experimental distributions of ZMW fluorescence inter-pulse durations sequencing in various regions of human chromosome 20 (circles) with the least squares fit by $P(\Delta\mu)$ (red lines).

**Figure 7. a)** Representative distributions of $\Delta\mu_i$ values computed from 150 kbase segments of human chromosome 1 (blue) and their least square fits by $P(\Delta\mu)$ (red lines). **b)** Inter-species variability of $\Delta\mu_i$ values, represented by parameters A and a of $P(\Delta\mu)$ obtained by least squares fitting of $\Delta\mu_i$ histograms computed from 150 kbase segments in respective genomes.

**Figure 8.** Comparison of $P(\Delta\mu)$ distributions computed in complete chromosomes (blue) and in coding regions (exome) only (black) in chromosomes 1, 2 and Y.

Systematic shift of exome distribution means to higher $\Delta\mu_i$ is due to the additional restriction of exon DNA sequences by amino acid coding.

**Figure 9.** Comparison of entromic coherence matrices in regions with high (top) and low (bottom) incorporation energy $\Delta\mu_i$ computed by exome co-processing for human and Plasmodium (**a,b**) and human and mouse (**c,d**). Human exome results are to the left from the arrow. ***R**$_{ij}$* intensities increase from magenta to red. (**e,f**) Comparison of $\Delta\mu_i$ distributions for the two species exomes.

**Figure 10**. Structural interpretation of shifts in the entromic coherence patterns for human (**a**) and mouse (**b**) polymerase β. The lines, connecting coherent regions in two species were computed by matrix alignment program (ref. 17). **c)** Projection of aligned coherence regions into human polymerase-substrate structure (3ISB).

**Figure 11.** Plots of ***M**$_i$* and $1/M_i$ profiles in exons of human CFTR gene. Only exons 8 and 13 have dominant $1/M_i$, indicating that corresponding highest incorporation entropy contributes to their specific disease-related splicing.

**Figure 12**. **a)** Example of difference $\Delta\mu_i - \Delta\mu_{i,wt}$ profiles for all synonymous mutants in one of the CFTR gene exon 13 position, studied in ref. 18. **b)** Linear relationship between the total $\Delta\mu_i - \Delta\mu_{i,wt}$, computed as integrated areas under the difference curves and the experimental percentage of spliced CFTR variant expression.

_________________________________________________________________________

**Figure S1.** Scheme explaining computation of incorporation entropy vectors from DNA sequence. Series of position-centered segments with systematically increasing length is selected from DNA sequence (top, open rectangles) and processed into incorporation entropy. Simultaneously, the computation is performed for segments of the same length in the center base of always identical "buffer" DNA sequence, which facilitates the common cyclic boundary conditions at the DNA ends (bottom, shaded rectangles). The incorporation entropy values from the reference computation are used for normalization as shown.